\documentclass[article,a4paper,twoside,twocolumn]{revtex4}
\usepackage{amsmath,amsfonts,graphicx}
\usepackage[utf8]{inputenc}
\usepackage[english]{babel}
\usepackage[T1]{fontenc}
\usepackage[amssymb]{SIunits}
\usepackage[active]{srcltx}
\usepackage{anysize}
\marginsize{2cm}{2cm}{1,5cm}{2cm}

\begin{document}
\bibliographystyle{apsrev}
\title{An intuition behind quantum measurement}
\author{Piotr Witas}
\email{pjw@fizyka.umk.pl, pjwitas@gmail.com}
\affiliation{Institute of Physics, Nicolaus Copernicus University,\\Grudziądzka 5/7, 87-100 Toruń, Poland}
\date{\today}
\begin{abstract}
An attempt is made to give a heuristic explanation of the distinguished role of measurement in the quantum theory. We question the notion of ``naive'' reductionism by stressing the difference between an isolated quantum and classical object. It is argued that the transition from the micro- to the macroscopic description should be made along some parameters not characterized by the quantum theory.
\end{abstract}
\maketitle

\section{Introduction}
One of the most serious problems of the quantum theory can be, in short, formulated as follows: its formalism is extremely accurate, but it is difficult to find solid conceptual foundations for what it actually describes. Perhaps the only useful attitude is to claim that it provides information exclusively on measurement outcomes. However, it turned out to be surprisingly hard to give the term ``measurement'' a clear meaning on physical grounds. A consequence is that it is still unknown how to obtain the classical ontology from this theory in the macroscopic limit.\\
\indent In this paper we would like to address some aspects of the relation between the quantum and classical domains. Our approach is definitely a heuristic one (basically, the present work could have been written right after the formulation of quantum mechanics in the 20's, since we do not invoke any intricate mechanisms described by the theory itself). We aim at characterising, in a very general and qualitative manner, a conceptual scheme which, on one hand, would be in agreement with the crucial empirical facts known from the two regimes, and on the other - that would reconcile the solid picture of the world known from the classical with the fuzziness that the quantum one seems to predict for the macroscopic level. However, we do not want to speculate on what ``actually happens'' in the quantum world - we only ask how to include a smooth deformation of classical intuition into our understanding of the world to make the quantum theory less in conflict with the classical one.\\
\indent In the following, we question very mildly the reductionistic attitude inherited from the XIX-century classical statistical mechanics. In other words, we cast doubt on the validity of an argument often encountered in works related to the interpretational problems of the quantum theory: ``after all, macroscopic objects are built of microscopic ones''. We do it by concentrating on the well-known fact that the quantum state describes inherently context-dependent entities, contrary to the classical one. It is suggested by these considerations that a deeper description than offered by the quantum theory would be more natural for the treatment of quantum-to-classical transition, and that it is much easier to interpret the theory as a one that lacks the description of an important piece of physics from the very beginning (i.e. as a phenomenology).\\
\indent We point that no ``complete'' explanation of the conceptual strangeness of the quantum theory is given here. Specifically, what quantum measurement really is remains unexplained. Rather, we find a place for the known quantum features in a larger framework.\\
\indent To make as close connection with the classical intuition as possible, we make much use the notion of object. At first, our construction may seem of restricted generality: we know that relativistic quantum physics strongly suggests to avoid using objects. However, our conclusions are valid for the observables themselves.

\section{Quantum peculiarities}
To set the stage, we give a brief survey of characteristic quantum features that we are going to address specifically.\\

\noindent\textbf{State vector collapse}. As is widely known, the following applies to the time evolution of an ``isolated quantum system'':
\begin{itemize}
\item between a preparation at $t=0$ and a measurement at $t=\tau$ its pure state, represented by a normalised vector from the Hilbert space, evolves according to
\begin{equation}
\label{unitary_evolution}
|\psi(\tau)\rangle=e^{-iH\tau}|\psi(0)\rangle,
\end{equation}
where $H$ is the system's Hamiltonian (we set the units so that the reduced Planck's constant $\hbar=1$);
\item after a projective measurement (we do not consider other cases in this work) of an observable with a non-degenerate spectrum, the state collapses to an eigenstate $|i\rangle$ of the measured observable $O$ giving the corresponding eigenvalue as a result, with probability dictated by the Born rule: $p_{i}=|\langle i|\psi\rangle|^{2}$.
\end{itemize}

\noindent The second law seems to be an intriguing mixture of a physical process and a change of knowledge. It is stochastic and instantaneous (at least on the level of a postulate), yet it cannot be a probability distribution as classical mixed states are, since it seems to create the values of observables. What is even more strange, the created properties can be only ``partly'' fixed, as seen, for instance, in the position measurement in non-relativistic quantum mechanics. The measurement ``checking'' that a particle ``is'' in a space volume $V$ reduces the support of the quantum state
\begin{equation}
|\psi\rangle=\int d^{3}\bold{x}\ \psi(\bold{x})|\bold{x}\rangle
\end{equation}
with the projector
\begin{equation}
\label{eq:loc}
\int_{V} d^{3}\bold{x} |\bold{x}\rangle\langle \bold{x}|,
\end{equation}
where $|\bold{x}\rangle$ constitute a basis of position operator eigenvectors. However, it is hard to speak about the position of the particle within the volume $V$ in the reduced state.
Last but not least - the collapse is not a unitary transformation, what makes it fundamentally different from the evolution in (\ref{unitary_evolution}). However, one would rather expect measurement to be an ordinary physical process.\\

\noindent\textbf{Macroscopic superpositions}. A peculiar feature of the quantum theory is a possibility of producing superpositions on a macroscopic scale. This is done through establishing quantum correlations between a microscopic quantum system and a large set of such entities. One typically visualizes this in the von Neumann measurement scheme \cite{vn} and the Schroedinger's cat paradox \cite{Schr} (see also \cite{Schloss} for a much more thorough description of the so-called measurement problem). Let us briefly describe their common features. Consider states $|s_{i}\rangle$ of a quantum system and states $|a_{j}\rangle$ of the macroscopic object (we accept the universality of the theory). For clarity, we assume that there exists a ``ready'' state, $|a_{0}\rangle$. Since we want to describe the process of establishing correlations quantum-mechanically, we suppose the existence of a unitary operator $U$ which does the following:
\begin{equation}
|s_{i}\rangle|a_{0}\rangle\overset{U}{\longrightarrow}|s_{i}\rangle|a_{i}\rangle
\end{equation}
for every $i$. From the linearity of $U$ it immediately follows that:
\begin{equation}
\label{eq:def_out}
\Big(\sum_{i}\alpha_{i}|s_{i}\rangle\Big)|a_{0}\rangle\overset{U}{\longrightarrow}\sum_{i}\alpha_{i}|s_{i}\rangle|a_{i}\rangle.
\end{equation}
The above equation describes a large, macroscopic system. However, since large-scale objects are never observed in superpositions, this seems to be a pathology. Put differently, it is not clear what a superposition should mean in the context of macroscopic bodies, since their properties in such states are not definite. In case of the von Neumann scheme, (\ref{eq:def_out}) leads to the so-called problem of definite outcomes.\\
\indent We stress that the above being possibly a description of measurement is of secondary importance to us. What really counts is that quantum superpositions can in principle be amplified to the macroscopic level through establishing quantum correlations.

\section{Reductionism and context-dependence}
A matter of fundamental importance for our picture of the world and the consistency of physical theories is how to relate micro- and macroscopic domains. It is \textit{\'a priori} acceptable that the microworld has its own, unintuitive laws, but nevertheless we would like to obtain a proper classical limit from the quantum theory, in particular concerning the ontology. Perhaps our starting point in establishing such relations would be to postulate that macroscopic objects we knew from our surroundings were simply \textit{built} of some microscopic ones. Such approach turned out to be very fruitful when classical statistical mechanics of gases and liquids was being formulated. There, the fundamental constituents of matter followed the same laws of motion as the macroscopic complex objects. Simplifying the issue a bit, we may say that the only significant difference between the small and the big objects was the size. Now, after quantum mechanics appeared on the scene, it was natural to consider it simply as a refined description of these elementary objects. However, quite soon the problems mentioned in the previous section started to appear - quantum systems of macroscopic size can be put into superpositions and properties of quantum objects are defined by the measurement procedure, what can hardly be called ``classical''. In such situation, a way out was to invent specific mechanisms that could produce at least the ``appearance'' of the classical world (employing the theory of decoherence, for instance) or somehow interpret the quantum theory keeping in mind the reductionistic attitude (see \cite{Au} for an exposition of the main approaches; this reference may be also used for all other quantum-foundational issues mentioned in this paper). Unfortunately, no general consensus was reached along these lines of research.\\
\indent At this point, a different - and perhaps a bit too radical at first sight - route may be taken. A conclusion that one might draw by examining the behaviour of elementary constituents of matter is that, \textit{treated naively}, the idea of building might simply be false: since the classical limit of the theory is so ill-defined, maybe the limiting procedure is applied to a wrong piece of our description of the world? In the author's opinion this is indeed so, and this is ultimately due the difference between classical and quantum isolation. Consider a single, isolated quantum point particle. It is evident from what has been recalled in the previous section that on the quantum level it no longer makes sense to talk about it as separated out of its environment (whatever this environment may be), contrary to the classical case, and characterized by properties like position, momentum, energy, angular momentum etc. Note that the state of such ``meaningless'' entity as our particle is described by a vector from the Hilbert space. Now, let us add some more (a lot more) quantum objects to the picture. What we end up with is again a vector in some Hilbert space. A question: why should this large compound object (treated as an isolated one) make more sense as a whole than a single isolated particle if we use the same formalism to describe them? We clearly see from this that an application of the quantum composition of elementary objects by building a large Hilbert space simply does not do the job of recovering the classical world, because the properties of a quantum object - no matter how large - are, in some sense, joint properties of this object and the ``measuring apparatus'', \textit{they are defined only in some context}. This context seems to be \textit{something external that quantum theory does not allow to purely express in terms of elementary quantum objects}. The whole picture still needs to be supplemented by some important ingredient - which is, of course, quietly introduced by the collapse postulate.\\
\indent The argument ``every child knows that every macroscopic body is \textit{built} of elementary quantum objects and should be describable by the quantum formalism'' can be immediately responded to with ``where do we know that from?''. ``Well, the theory, which always gives perfect predictions, says so'' is certainly not the right answer, looking at the problems we encounter in constructing the classical limit of the quantum theory. Thus, a remark concerning reductionism - we should not presuppose the validity of certain concepts, but rather infer it from the working piece of knowledge. In fact, what is implied by quantum physics is that we can \textit{find} some smaller objects inside bigger ones (we see them when we look for them) - and nothing more.\\
\indent Of course, reductionism in terms of quantum objects is essential for explaining many properties macroscopic bodies. In this sense, reductionism has perhaps never failed. It is thus obvious that it should be treated as one of the pillars of modern physics. But how can we then reconcile these two seemingly contradictory points? To remain consistent, we deduce that only a deviation from reductionism which is \textit{small} (i.e. unimportant from the operational point of view) is present in the quantum theory, and that only this part is responsible for conceptual problems we encounter in the case of quantum-to-classical transition. This is what makes us slightly modify the canonical view and instead of saying ``macroscopic objects have some collective features because they are sets of elementary quantum entities'' say ``macroscopic objects have some <<collective>> features because elementary quantum entities (or sets of them) can be somehow produced inside them by what we call measurement and what we do not understand''. We stress that these two should be clearly distinguished because the reductionism itself does not explain the classical ontology - a separate measurement postulate apparently has to be included \footnote{As a side remark - an obvious and well-known departure from reductionism is implied by quantum entanglement, which prohibits subsystems of a larger system to have individual, definite states.}.\\
\indent A natural question arises: what can we propose instead of the naive reductionistic approach that would include the mentioned deviation in a desired way? Let us for a moment shift our attention from objects to observables (we will be able to make conclusions on the former if we treat them simply as sets of properties). The idea now is to take the quantum-mechanical description of measurement seriously and assume that properties of quantum systems are indeed created - and to a different ``degree'', in accordance with the discussion around (\ref{eq:loc}) - during this process, whatever it is. We do not propose any underlying mechanisms allowing to reconstruct these quantities, but try to speculate what general changes we could introduce to our picture of the world by considering such possibility.\\
\indent To summarize what we have already said: whatever quantum theory describes, it does it very well. Our job now is to make the best conceptual fit for the formalism. Even if it explicitly points that the formalism is not fundamental.

\section{Observables as emergent quantities}
\label{sec:obs_emer}
Let us then stop treating known physical quantities (both familiar from the classical level, like position and momentum, and specifically quantum-mechanical ones, like spin) as fundamental. In other words, we propose that the range of validity of these quantities is in general not infinite \footnote{The plausibility of this is strengthened by the fact that the programme of \textit{hiding} the classical physical variables has not been conceptually too successful so far.}. In this context, let us introduce an entity (call it $\mathcal{E}$), from which they are \textit{emergent}. Strictly speaking, we treat this structure for the moment as the \textit{world}, and all of our physics as describing some emergent properties of it (in section \ref{sec:emer_from_wh} we point that it does not do much harm if we forget it, but it is much more convenient to keep it along the argumentation we present). Actually, we would like to say as little as possible about $\mathcal{E}$, not to hypothesize too much and to try to give a better conceptual foundation of the quantum theory without formally extending it. We only assume that the behaviour of this entity can produce at least two kinds of conditions:
\begin{itemize}
\item ones in which this emergence is clearly dynamical (we then refer to the \textit{dynamical level} of $\mathcal{E}$),
\item and ones in which the mentioned physical quantities are ``stable'' (then we talk about the \textit{stable level}; our experience tells us that this happens for instance on the classical level which lives its own life and is insensitive to fluctuations ``below''; let us assume, however, that the ``dynamical'' conditions may also appear on the macroscopic level - when we superpose a large collection of quantum objects).
\end{itemize}
In other words, we locate the micro- and macrolevel known from the experience somewhere in $\mathcal{E}$. We point that a given physical quantity, like position, refers to rather different entities in the two cases - on the microlevel it might be an elementary particle, on the macrolevel - a point (or rather what we see as a point) of a macroscopic object. The former is a ``partly determined'' physical quantity, as discussed in the vicinity of (\ref{eq:loc}). A position of some tiny region of a macroscopic body is thus not treated as \textit{really} a position of some set of elementary constituents. We make this distinction clearly since the big objects are no longer treated as built so directly from the small ones.\\
\indent Naturally, the scale is defined in this scheme by the characteristics of $\mathcal{E}$ (we may think of some ``ghost parameters'' that move us between the two levels described above). A consequence is that what serves as the definition of the scale in a naively reductionistic approach (concentration of objects, their size etc.) is now somehow only a product. That is, \textit{the existence of the stable level is established not by employing the quantum law of composition of some elementary objects, but by some completely unknown quantities}. Of course, one still can find more atoms in a cubic meter of a given substance than in a cubic nanometer of it, but the quantum-to-classical transition is governed by unknown to us characteristics of $\mathcal{E}$. \\
\indent It is worth stressing that since the familiar physical quantities usually do not exist on the dynamical level, almost everything we say about these quantities is related to the stable (and thus usually classical) one. First, this justifies treating the macroscopically visible parts of measurement apparatuses as purely classical in our approach (\textit{\'a la} orthodox interpretation). Second, it ``explains'' why we often need a classical theory to quantize, not the other way round - quantum theory serves in this approach to tell us in part about quantities which are emergent and present usually on the classical level.\\
\indent Observe that in such scenario, measurement may be looked upon as a dynamical process, describable only in terms of the dynamics of $\mathcal{E}$, and leading to a dynamical creation of the value an observable. Why should we be fond of that? There are at least a few good reasons:
\begin{itemize}
\item as mentioned, the values of observables really seem to be created, not discovered in this process;
\item the uncertainty principle becomes more comprehensible - the ``measured'' quantities are now indeed defined by the measuring procedure which depends on the details of the behaviour of $\mathcal{E}$; thus, if it is impossible to measure simultaneously two quantities, they cannot exist simultaneously and this would explain why so-called quantum particles do not have trajectories (by the way, this shows why observables are a bit more fundamental than objects, since we can have momentum without having position);
\item contextuality (in the sense of Kochen and Specker) also seems more ``acceptable'', since two different measurement procedures, having a common observable as their ``target'', may be related to essentially different dynamics of $\mathcal{E}$;
\item quantum Zeno effect also finds its ``justification'' - a state gets ``frozen'' and the quantum system cannot evolve when repeatedly measured because such continuous process can be viewed as sustaining the same dynamical state of $\mathcal{E}$, in which the values of observables have been produced, for some period of time.
\end{itemize}
Of course, the sole act of measuring is not the only phenomenon that is capable of creating the values of observables - the classical world should arise and operate even if there is not anyone to make measurements. We postulate that the dynamics of $\mathcal{E}$ leading to this creation simply occurs in what we usually take as measurement situations.\\
\indent Let us now come back to the notion of object. How does the above scenario relate to the observation that objects we see in experiments have their properties defined only in some context? The context-dependence means that it makes sense only to describe some larger set such objects, already equipped with properties, are a part of. As entities characterized by some \footnote{Take again a single non-relativistic particle. Note that the properties over which we superpose here - position, momentum, angular momentum, energy etc. - are relational. Indeed, there are ``absolute'' quantities describing a particle, like mass and electric charge, over which we cannot form superpositions, because well-known superselection rules forbid that (for mass, the restriction is known as Bargmann superselection rule, and it comes up in the context of the representations of the Galilei group). Thus the context-dependence may be reduced to relationalism of the mentioned quantities.} familiar physical properties, they simply do not exist beyond these sets. This in turn means that they are inconceivable alone. Thus, \textit{these objects have to arise together with their properties}.  Actually, we should reformulate a bit our observation that quantum objects are context-dependent. That is, if we understand such objects merely as collections of their properties, then, strictly speaking, such objects do not exist beyond the stable level.\\
\indent In summary, quantum objects do not have a character of some elementary entities or building blocks, but rather of a \textit{behaviour} of some underlying ``material''. That is the reason why we introduced $\mathcal{E}$. All this means that modern physics is about:
\begin{itemize}
\item emergent structures containing what we know as objects characterized by familiar physical quantities - this is the more or less well-defined classical region,
\item the emergence of these structures without knowing how it happens - this is the quantum part.
\end{itemize}
It also means that what we customarily call \textit{a quantum isolated object is a fundamentally different concept than a classical isolated one}. Classically (and thus macroscopically in practice), something is isolated if the influence of the surrounding objects on it is negligible. On the quantum level, however, the isolation means also depriving the object of its properties. If we do not allow for any other physical properties of such object, then ``quantum isolation'' means only an isolation of a particular Hilbert space form other Hilbert spaces, without reference to ``physical objects''.

\section{Quantum state}
It would now be in order to say how we understand the quantum state in the light of what has been said above. First of all, recall that the variables on which state depends are, on the physical side, emergent. This means that we need both a ``source'' and a ``detector'' to obtain meaningful numbers from the theory (in this case - probabilities), because only these two apparatuses taken together seem to let us stay on the level where the studied physical quantities are determined. The quantum state can then be safely regarded as only a means of relating different probability distributions, not as something describing ``a particle'' or whatever (usability of a single Hilbert space should not necessarily be interpreted as implying the existence of a single physical object; this implication ought to be true only relative to the stable level - where stochastic acts of creation of properties may be combined into a mental picture of a particle). It is only a mapping defined on position, momentum or directional (as for spin) space, not something corresponding to intuitively understood possibilities. That is, an expression like
\begin{equation}
\label{eq:q_state}
|\psi\rangle=\alpha|\bold{x}\rangle+\beta|\bold{y}\rangle
\end{equation}
should be taken as ``we have a mapping defined on two points in space'', not as ``there is a particle smeared over the region consisting of two points''. Conceptual problems seem to start appearing if we make the following association:
\begin{equation}
|\bold{x}\rangle\leftrightarrow \bold{x}.
\end{equation}
This, however, should never be done, since the former (which is quantum) means that there is a chance that some physical position described by co-ordinates $\bold{x}$ will be created, whereas the latter (which is classical) describes a physically existing position at $\bold{x}$. It is important to note that this is always true, even in the case when a system is in an eigenstate.\\
\indent Actually, in the context of previous sections it would be more appropriate to consider exclusively the probability distributions and forget about attributing any meaning to the state. We would then have only a single law of time evolution - concerning solely the probabilities. This, however, would be a bit inconvenient, since the state allows us to include the conditioning of these distributions on what happens on the stable level in a very natural way (this conditioning is done, of course, through the infamous collapse; the question if it is really instantaneous does not seem to be easy to answer, but at least we see that the issue \textit{should} be ambiguous - after all, we now nothing about the dynamics of $\mathcal{E}$).\\
\indent We would now like to comment on a very important issue concerning the quantum-to-classical transition, i.e. the implications of what goes under the name ``theory of decoherence''. It is interesting to see how one can relate the classical-like features it produces (including decoherence proper and selection of pointer states by environment; see \cite{Schloss} for a discussion and a list of important references) with our approach. For that purpose, imagine that $\mathcal{E}$ produces two classical structures $\sigma_{1}$ and $\sigma_{2}$ (on the stable level) that are isolated from one another. That is, within them the quantities like position and momentum exist, but they do not relate the structures themselves. We can model the scenario known from the Schroedinger's cat paradox with these - $\sigma_{1}$ may correspond to the interior of the box, $\sigma_{2}$ to the exterior containing the observer. Now, assume that, as a result of the dynamics of $\mathcal{E}$, $\sigma_{1}$ and $\sigma_{2}$ merge. From the point of view, say, $\sigma_{2}$, the structure $\sigma_{1}$ gets created in it (and vice versa). This means that many observables, pertaining to many objects, get their values fixed in $\sigma_{2}$ at the same time. \textit{Since we expect classical-like correlations between them, the quantum formalism should be able to mirror these}. Besides that, it is interesting to note that the ``border'' between these structures, obviously going through $\mathcal{E}$, would naturally correspond to the so-called Heisenberg cut.

\section{Emergent from what?}
\label{sec:emer_from_wh}
The reason why we introduced $\mathcal{E}$ was to make ``more room'', so as to - in our opinion - more naturally redistribute certain conceptual elements of the quantum theory, to embed its conceptual structure in a wider one, so as to make it more comprehensible. Obviously - we now nothing more about $\mathcal{E}$, at least at the moment. We think that discovering $\mathcal{E}$ essentially amounts to discovering new physics, explaining, for instance, the dynamical formation of quantities like position, momentum etc.\\
\indent However, one may look at the situation from a bit different angle. We may simply say that the precise knowledge of $\mathcal{E}$ is unimportant here - we only want the relations between different parts of the quantum formalism to be established. We are convinced that this whole presentation makes it plausible that \textit{the quantum theory is indeed a description of a world in which the precise shape of $\mathcal{E}$ has been forgotten}. In particular, it goes together with two trademarks of the quantum theory: a lack of ontology (which is obviously realised by $\mathcal{E}$ now) and a distinguished role of the observer (who has to check by his own eyes what happens in the laboratory since he does not know the precise dynamics of $\mathcal{E}$). \textit{A very important note is that this may be applied even if it is for some reason impossible for us to discover a better characterization of $\mathcal{E}$}. Of course, it is now the primary goal to look for such structure as $\mathcal{E}$, but this is not needed to give a qualitative interpretation of the quantum theory itself.

\section{Conclusions}
We have argued that perhaps the most serious conceptual problem of the quantum theory - related to the prediction of macroscopic superpositions - results from sticking to a straightforwardly understood notion of reductionism. By using the intuitive notions of object, context-dependence and emergence we have shown that quantum theory looks \textit{as if} there existed an unknown entity which literally produces the quantities we call ``observables'' in the process we customarily call ``quantum measurement''. We argued essentially that the physical principle behind the fundamental conceptual difference between quantum and classical mechanics is, metaphorically speaking, the following: when we describe \textit{physical object}, we cannot do that without describing other \textit{physical objects}. To be more precise - one cannot describe a particle with its position without mentioning other things that also have a position (measurement apparatus, for instance). In other words, in the quantum case one cannot introduce a kind of a background on which the objects are living (phase space, for instance). However, the process of including more and more entities should not be conducted on the level of the quantum formalism.\\
\indent Obviously, we have not said anything about a formal derivation of the theory based on the introduced conceptual scheme. This issue is going to be addressed in forthcoming papers.

\end{document}